\documentstyle[bo99,epsfig]{article}

\title{Gamma-loud quasars: a view with BeppoSAX }
\author{Tavecchio F.$^1$, Maraschi L.$^1$ Ghisellini G.$^1$, 
Celotti A.$^2$, Chiappetti L.$^3$
Comastri A.$^4$, Fossati G.$^5$, Grandi P.$^6$, Haardt F.$^7$, Pian E.$^8$, 
Tagliaferri G.$^1$, Treves A.$^7$, Raiteri C.M.$^9$, Sambruna R.$^{10}$, 
Villata M.$^9$}
\affil{
$^1$Osservatorio Astronomico di Brera, Milano, Italy,
 $^2$SISSA/ISAS, Trieste, Italy,
 $^3$IFC/CNR, Milano, Italy, 
 $^4$Osservatorio Astronomico di Bologna, Bologna, Italy,
 $^5$CASS/UCSD, La Jolla, USA,
 $^6$IAS/CNR, Roma, Italy,
 $^7$Universita' dell'Insubria, Como, Italy, 
 $^8$ITESRE/CNR, Bologna, Italy,
 $^9$Osservatorio Astronomico di Torino, Pino Torinese, Italy,
 $^{10}$PennState University, USA}

\begin{document}

\maketitle

\begin{abstract}
We present $Beppo$SAX observations of the $\gamma $-ray emitting 
quasars 0836+710, 1510-089
and 2230+114. All the objects have been detected in the PDS up to 100 keV and
 have
extremely flat power-law spectra above 2 keV ($\alpha _x$=0.3--0.5). 0836+710 
shows absorption higher than the galactic value
and marginal evidence for the presence of the redshifted 6.4 keV Iron line. 
1510-089 shows a spectral break
around 1 keV, with the low energy spectrum steeper ($\alpha _l$=1.6) than the high energy 
power-law ($\alpha _h$=0.3). 
The data are discussed in the light of current Inverse Compton models for the
 high energy emission.
\end{abstract}

\section{Introduction}
Since the EGRET detection of about 60 blazars as strong $\gamma $-ray emitters
the study of these extreme objects has received a renewed interest.
The overall Spectral Energy Distribution (SED) of Blazars shows two broad components, the first one
peaking at IR-up to soft X-rays, the second one in the $\gamma $-rays, from MeV 
up to TeV energies.
The first peak is due to synchrotron radiation produced by relativistic
electrons, while the high energy component is believed to be Inverse Compton
scattered radiation. The seed photons for the IC scattering could be the
synchrotron photons themselves (SSC model) or photons produced in the region external to the
jet (EC model) which, especially for quasars with strong emission lines, 
is probably rich of optical-UV radiation.
The subclass of quasar-like sources contains the most luminous sources,
with apparent $\gamma $-ray luminosity up to $10^{48}$ erg s$^{-1}$. 

In the following we present the $Beppo$SAX observations of three gamma-loud quasars (0836+710,
2230+114 and 1510-089), detected up to 100 keV with the high energy
instrument PDS, and we discuss the external Compton scenario. A full
paper is in preparation (Tavecchio et al. 2000).

\section{The observed objects}

\noindent
$\bullet$0836+710: this is a distant quasar ($z=2.172$), characterized
by a very flat X-ray spectrum, observed with ROSAT and ASCA. The ASCA observation
showed a column density greater than the galactic value (Cappi et al. 1997).

\noindent
$\bullet$2230+114: this source ($z=1.037$), observed with GINGA,
ROSAT and ASCA (Lawson \& Turner 1997, Brinkmann et al. 1994, Kubo et
al. 1998)
shows a flat spectrum extending smoothly in the gamma-ray band, as indicated
by OSSE observations (Mc Naron-Brown et al. 1995).

\noindent
$\bullet$1510-089: this interesting Highly Polarized Quasar ($z=0.361$) shows 
a pronounced
UV bump (Pian \& Treves 1993). The EXOSAT observation (Singh et al. 1990) 
suggested the presence of a fluorescence iron line, not present in a more 
recent ASCA observation (Singh et al. 1997).

\begin{center}
\begin{table}
\begin{tabular}{cccccc}
\hline
\hline
$\Gamma $ & \, \, $E_b^*$ & \, \, $\Gamma _h^* $\, \, & $N_H$ & 
$F_{[2-10\, keV]}$ &  $\chi^2/$d.o.f.\\
&\, \, keV & & 10$^{20}$ cm$^{-2}$ & $10^{-12}$ erg cm$^{-2}$ s$^{-1}$ &  \\
\hline
\multicolumn{6}{c}{ {\bf 0836+710}} \\ 
\hline
$1.32\pm 0.04 $ & -& -& $8.3^{13.0}_{5.7}$ & 26 & 63.47/63\\
$0.83^{1.17}_{0.3}$ & $1.2\pm 0.3$& $1.31\pm 0.03$& 2.98(fix)& 26& 63.14/62\\
\hline
\multicolumn{6}{c}{ {\bf 2230+114}} \\ 
\hline
$1.51\pm 0.04$& - &- & $7.3^{11.1}_{4.6}$ & 6.05 & 51.11/51 \\
\hline
\multicolumn{6}{c}{ {\bf 1510-089}} \\ 
\hline
$1.35 \pm 0.07 $& -& -&  0-2.44 & 5.3 & 60.24/64\\
$2.65^{3.28}_{2.05} $& $1.3\pm 0.3$ & $1.39 \pm  0.08$ & 7.8 (fix) & 5.3 & 43.06/63 \\
\hline
\multicolumn{6}{l}{$^*$: only for the broken power-law model} \\
\end{tabular}
\caption{Fits to {\it Beppo}SAX Data (LECS+MECS+PDS).}
\end{table}
\end{center}

\section{ $Beppo$SAX Observations and results}

We modelled the spectral data with either single or broken power-law models (with 
galactic and free absorption). Results of the spectral fits to 
the LECS+MECS+PDS data are shown in Table 1. In the following we discuss
the results for each object.

\noindent
{\bf 0836+710}: the data are consistent both with a power-law model with 
intrinsicabsorption ($N_H\simeq 7\times 10^{21}$ cm$^{-2}$ in the QSO rest 
frame) and with  a broken power-law model with fixed galactic absorption.
With both models the residuals show an excess at about 2 keV, that could be 
interpreted as the redshifted fluorescence iron line (see Fig 1). Adding a gaussian line
with energy as a free parameter the fit converges to an energy of $E=1.99 \pm 0.1$ with
an intrinsic equivalent width of $EW\simeq 110$ eV, but the significance of the 
improvement of the $\chi ^2$, evaluated with the F-test, is not very high 
($P \simeq 90\%$).

\noindent
{\bf 2230+114}: A simple power-law model ($\alpha =0.5$) with absorption consistent with the galactic value 
reproduces quite well the data in the whole range 0.1-100 keV.

\noindent
{\bf 1510-089}: A simple absorbed power-law (although statistically acceptable, see 
Table 1) gives evident residuals at low energies. The $F$-test
confirms that a broken power-law is a better model (with a probability 
$P>99.99 \%$). The PDS/MECS relative normalization (1.3-2.5 90\% conf. level) 
is above the accepted range (0.77-0.93); this problem is possibly due to
contamination by a source in the large FOV of PDS ($\sim 1$ deg.). In the fit
we fixed the normalization to 0.85.

\begin{center} 
\begin{table}
\begin{tabular}{cccccccc}
\hline
\hline
$R_{16}$ & B& $\delta$ & $\gamma _{min}$ & n & $L_{inj,45}$ & $\tau L_{ext,45}$ &$R_{ext,18}$\\
(cm)& (G)& & & &(erg s$^{-1}$) &(erg s$^{-1}$) & (cm) \\ \hline
\multicolumn{8}{c}{\bf 0836+710} \\
\hline
4 & 5.3 & 18 & 50 & 3.05 & 14.8 & 3.2 & 1.5 \\
\multicolumn{8}{c}{\bf 2230+114} \\
\hline
4 & 3.7 & 15.5 & 130 & 3 & 0.14 & 0.4 & 1 \\
\multicolumn{8}{c}{\bf 1510-089} \\
\hline
2 & 3.1 & 17 & 65 & 2.9 & 0.01 & 0.45 & 1.\\
\hline
\end{tabular}
\caption{Parameters used for the EC model}
\end{table}
\end{center}

\section{Discussion}

\noindent
$\bullet$ We constructed the SEDs of the observed sources using contemporaneous
X-ray and optical observations and historical data taken from the literature 
(an example is reported in Fig.1)
We have reproduced the observed spectrum using the homogeneous
EC model discussed in detail in Ghisellini et al. (1998). In a spherical region
with size $R$, a power-law electron distribution (with slope $n$ and limits
$\gamma _{min}$ and $\gamma _{max}$) is continuously injected with luminosity
$L_{inj}$. Electrons cool
through the synchrotron and IC processes and are free to escape from the 
source at some velocity $v_{esc}$, forming a flat ($\alpha < 0.5$) power-law below $\gamma _{min}$. 
We assume that the external radiation field is described by a black body 
spectrum with luminosity $L_{ext}$, diluted in a spherical 
region with size $R_{ext}$. The parameter values for the models are reported 
in Tab. 2.

\noindent
$\bullet$ It is interesting to note that the derived spectral indices 
in the medium to hard X-ray band are flatter than 0.5 in two out of three
sources. A population of electrons cooling through synchrotron and IC forms
a distribution which produce a spectrum with $\alpha =0.5$. Therefore 
some additional mechanism is required in order to produce the 
observed flatter spectrum, e.g. escape or injection of an intrinsically
flat distribution (see e.g. Ghisellini 1996).

\noindent
$\bullet$ The excess of 1510-089 in the soft band is well understood 
as due to the SSC emission (see Fig.1), although another possible
source is the tail of the strong UV bump.

\noindent
$\bullet$ We confirm the presence of absorption higher than the galactic one in 0836+710. 
The origin of this absorption is likely intrinsic to the source. The
fluorescence iron line suggested by our data could
be produced through reprocessing by the same material responsible for the 
absorption. On the other hand a broken power-law continuum with galactic absorption can
reproduce the data equally well: in this case the break
could be due to the incomplete comptonization of the soft external photons
(see Ghisellini 1996).

\begin{figure}
\hskip -1.1cm
\vskip -1.9cm
\psfig{file=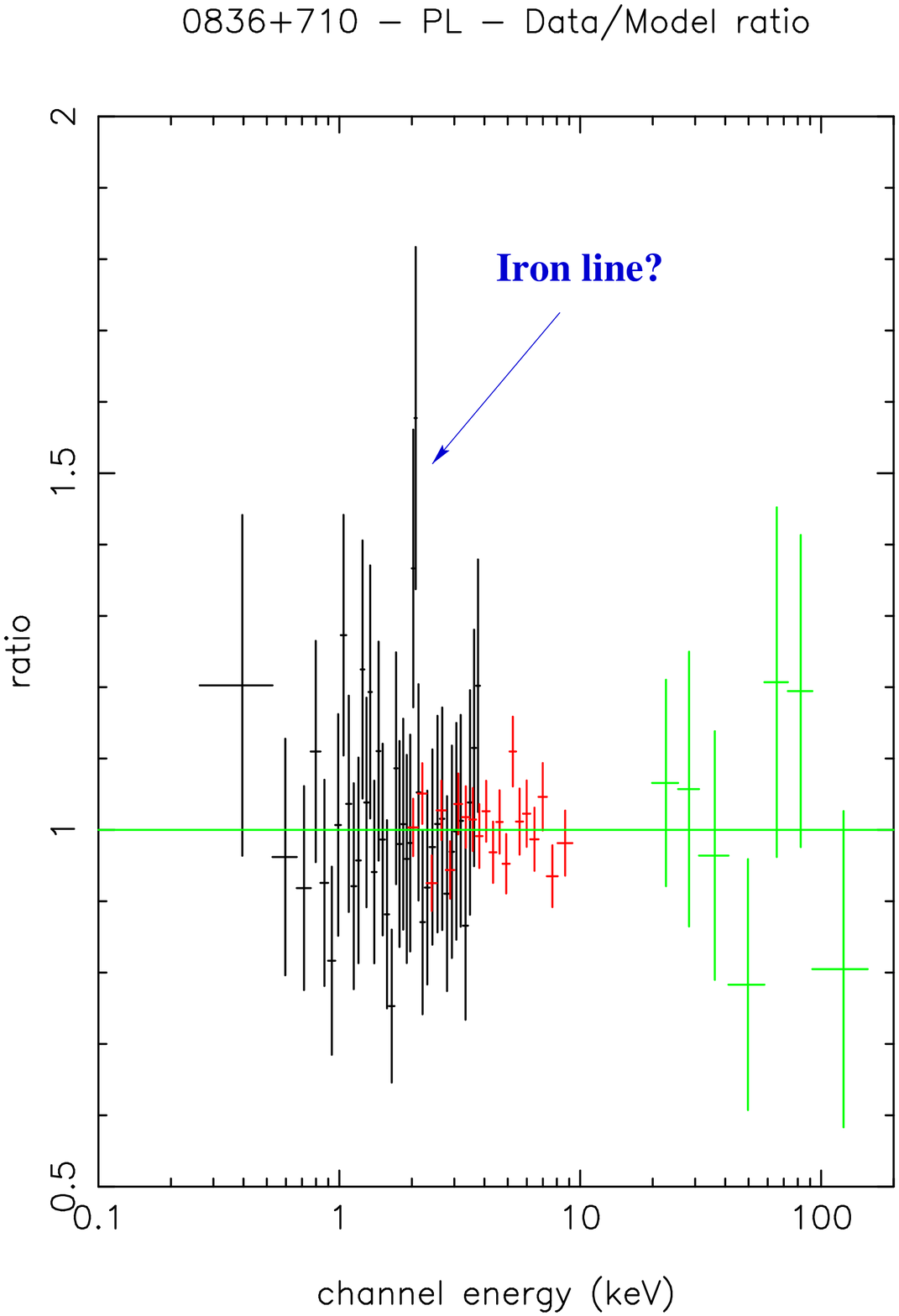, width=5.7cm, height=6cm}
\vskip -6.4cm
\hskip 5.35cm
\psfig{file=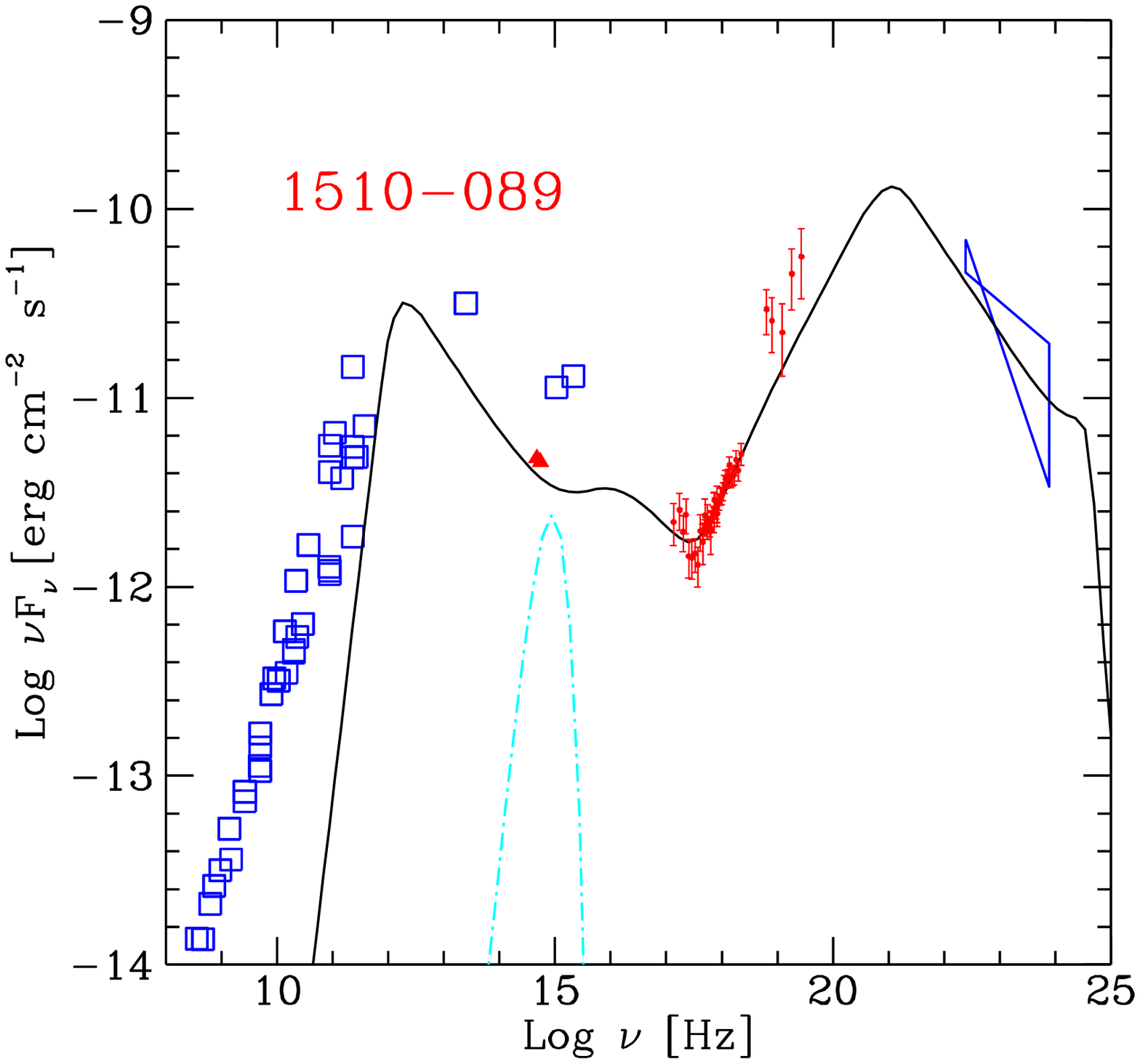, width=7.cm}
\caption[]{Left panel: Data/Model ratio for 0836+710 with the power-law model 
($N_H$ free). Right panel: Overall SED of 1510-089 with the spectrum 
calculated using the homogeneous EC model. Radio to UV (open squares) and 
gamma-ray data are taken from the literature.}
\vskip -.2cm
\end{figure}














\begin{references}
\ref Brinkmann, W. \& Siebert, J. 1994, A\&A, 285, 812

\ref Cappi, M., et al. 1997, ApJ, 478, 49

\ref Ghisellini, G. 1996, IAU Symposia, 175, 413

\ref Ghisellini, G., et al.1998, MNRAS, 301, 451

\ref Kubo, H., et al. 1998, ApJ, 504, 693

\ref Lawson, A. J. \& Turner, M. 1997, MNRAS, 288, 920

\ref McNaron-Brown, K., et al. 1995, ApJ, 451, 575

\ref Pian, E. \& Treves, A. 1993, ApJ, 416, 130

\ref Singh, K. P., et al. 1990, ApJ, 365, 455

\ref Singh, K. P., et al. 1997, ApJ, 491, 515 

\end{references}
\end{document}